\documentclass[manuscript]{aastex61}
\usepackage{graphicx}
\usepackage{natbib}
\usepackage{amsmath}
\usepackage{hyperref}
\usepackage{bm}

\newcommand{\upd}{\mathrm{d}}
\newcommand{\2}{$_2$}

\shorttitle{}

\begin{document}

\title{Global or local pure-condensible atmospheres: importance of horizontal latent heat transport}



\author[0000-0001-7758-4110]{F. Ding}
\affil{Department of the Geophysical Sciences, University of Chicago, Chicago IL 60637, USA}
\email{fding@uchicago.edu}

\author[0000-0002-5887-1197]{R. T. Pierrehumbert}
\affil{Department of Physics, University of Oxford, Oxford OX1 3PU, UK }


\begin{abstract} 
The distribution of a pure condensible planetary atmosphere in equilibrium with a surface reservoir is revisited employing the energy budget of the climate system, emphasizing the atmospheric horizontal latent heat transport. This configuration is applicable to icy
Solar System bodies such as Triton as well as a range of possible exoplanet atmospheres,
including water or $\mathrm{CO_2}$ iceballs or ocean worlds, and lava planets with mineral vapor atmospheres. Climate regimes for slowly rotating planets with the hot-spot near the substellar point, and for rapidly rotating planets with a warm equatorial belt, are both treated. 
A non-dimensional parameter controlling the fractional variation of the surface pressure is derived;  it measures whether the pure condensible atmosphere is global or localized \deleted{near the substellar point}.  The global pure condensible atmosphere with the non-dimensional parameter much less than order of unity is maintained by the strong horizontal latent heat transport associated with \replaced{the ``evaporation-driven flow''}{an ``evaporation/sublimation-driven flow''} from warm to cold places that \replaced{flattens}{compensates for} the incoming differential radiative forcing. We show that the variation of surface temperature can be estimated in terms of this non-dimensional parameter if it is not too large. \deleted{One implication of the result is that low-spin planets with water dominated atmospheres should have globally uniform surface temperature, leading to Snowball or Waterworld states but not Eyeball states.} \added{In the case of a pure water-vapor atmosphere with an ice or liquid surface, we show that the atmosphere is thick enough to maintain nearly isothermal surface conditions even when the substellar surface temperature is around the freezing point.} Finally, it is proposed that the \replaced{evaporation-driven}{evaporation/sublimation-driven} flow regime for global atmospheres could be detected via its effect on the inhomogeneous distribution of minor non-condensible components in the atmosphere. 

\end{abstract}

%
%



%
%
 
\section{Introduction}\label{sec:intro}

Condensation is ubiquitous in planetary atmospheres, and forms various kinds of condensates that play an important role in the atmospheric radiative, chemical and dynamical processes. The most familiar case is $\mathrm{H_2O}$ condensation that is prevalent on  Earth and the gas giants. Other examples includes CH$_4$ on Titan, NH$_3$ and NH$_4$SH on Jupiter and Saturn, and CH$_4$ on Uranus and Neptune. As is the case for water vapor in the Earth's atmosphere, the condensible substances in these atmospheres are dilute \added{in the sense that the condensible component makes up a small fraction of the atmosphere.} \replaced{, and therefore can}{These cases can therefore } be treated using methods familiar from the extensively studied case of moist convection on the present Earth. \added{The distinction between the dilute and nondilute behavior of atmospheres with a condensible component is discussed at length in \citet{pierrehumbert_dynamics_2016}. }

\replaced{Observations suggest on some of icy satellites in the Solar system the atmosphere is dominated by condensible components and most of the planetary surface covered by the condensate is in vapor equilibrium with the above air (e.g., Io with condensing SO\2 atmosphere and Triton with condensing N\2 atmosphere).}{Observations of some of the icy satellites in the Solar system suggest that the atmospheres are dominated by condensible components and most of the planetary surfaces covered by the condensate are in vapor equilibrium with the air overlying the surface (e.g., Io with condensing SO\2 atmosphere and Triton with condensing N\2 atmosphere). } This kind of atmosphere, referred to as pure condensible atmospheres in this paper, has a completely different circulation regime compared to atmospheres with the dilute condensible component, as a consequence of the strong connection between the surface temperature and pressure via the Clausius-Clapeyron relation. Pure condensible atmospheres were also proposed to form under various scenarios with different kinds of volatiles, e.g.,  a dense condensing CO\2 atmosphere on Early Mars \citep{forget_3d_2013} and on planets near the outer edge of the habitable zone \citep{wordsworth_is_2010}, condensing N\2 atmosphere on Pluto \citep{forget2017Pluto} and on Early Titan \citep{charnay_titans_2014}, and the extreme hot rocky vapor atmosphere on close-in rocky exoplanets \citep{castan_atmospheres_2011}. In all these cases, the atmosphere is supplied by vaporization \added{or sublimation} of a volatile reservoir near the warmest part of the planet, and flows to its sink where it condenses in the colder regions (e.g. the nightside of a tide-locked planet). 

It has long been recognized that pure condensible atmospheres could be either global -- with fractional surface pressure variation much less than order of unity as for N\2 on Triton -- or local near the source regions as for SO\2 on Io. \replaced{The fractional variation of surface pressure has been linked to the thickness of the atmosphere}{The thickness of the atmosphere has been implicated in determining whether the atmospheric circulation can carry enough heat to keep the surface pressure variations small } \citep{trafton_global_1983, trafton_large_1984}. In this paper, we revisit this problem by means of the horizontal atmospheric energy transport equation with emphasis on the latent heat transport. A non-dimensional parameter that determines the variation of surface pressure is derived from the energy budget of the pure condensible atmosphere. Idealized models are developed for two limiting cases: (1) rapid rotators for which the radiation balance can be zonally averaged, the dominant temperature gradient is from the hot equator to the cold pole, and Coriolis forces dominate the dynamics, and (2) slow rotators for which the dominant temperature gradient is from the hot substellar point to the cold antistellar point\deleted{, the circulation is axisymetric about the line passing from the substellar to antistellar point, and Coriolis forces are negligible}.  Tide-locked planets in sufficiently long period orbits are examples of the latter regime, but bodies with weak thermal inertia that have nonzero rotation relative to the substellar point, such as Triton or Pluto, can also be in this regime. 
The simplified solution helps to broaden the understanding of the range of possible atmospheric configurations for pure condensible atmospheres on extrasolar rocky or icy planets, including the intriguing case of water-dominated atmospheres over an N\2 and CO\2-poor Snowball or Waterworld.  

\added{
The thermodynamic and radiative assumptions on which our model rests are common to both classes of spin states, and are summarized as follows:
\begin{itemize}
   \item The atmosphere consists of a single chemical substance, which is condensible in the range of temperatures encountered at the surface of the planet.  It is assumed that the surface pressure is everywhere equal to the saturation vapor pressure corresponding to the local surface temperature. This assumption is met in the most straightforward way if there is a global surface reservoir, but that restriction could be relaxed in places where the atmosphere is cold enough to be condensing out onto the surface. 
    \item The atmosphere is in hydrostatic balance. This is important, because in hydrostatic balance the surface pressure determines the mass contained in a column of the atmosphere. 
    \item Latent heat dominates the energy transport.
    \item The infrared cooling to space (i.e, the outgoing longwave radiation, $OLR$) can be written as a function of the
          surface temperature.
\end{itemize}
The dynamic assumptions differ between the two spin cases considered.  For slow rotators, it is assumed that all atmospheric properties are axisymmetric about the axis joining the substellar to the antistellar point, and that Coriolis effects are negligible. It is also assumed that the vertically integrated mass transport can be estimated in terms of the surface wind. These assumptions are the same as those made in \citet{ingersoll_supersonic_1985}.  In the case of fast rotators, the assumption is that the atmosphere is symmetric about the spin axis, and that the mass and energy transports are dominated by the Ekman transport, without any significant contribution from transient eddy effects.  The circumstances under which the various assumptions are justified will be discussed in the course of the exposition below.
}

\section{Energy budget on planets with pure-condensible atmospheres}
 
A pure condensible atmosphere is an atmosphere consisting of a single substance (e.g. $\mathrm{N_2}$ or $\mathrm{H_2O}$) which can condense over the range of temperatures encountered in the atmosphere. \replaced{The model we shall formulate applies to pure condensible atmospheres which are saturated everywhere (i.e. that the vapor pressure equals the saturation vapor pressure), though subsaturation in a high-altitud stratosphere would do little to change the results. For a pure condensible atmosphere, saturation is maintained where a surface reservoir is evaporating or sublimating because any subsaturation would lead to a large near-discontinuity in pressure at the surface, which would drive strong mass flux into the atmosphere. Away from the atmospheric source region, where the cooling atmosphere condenses onto the surface, saturation is also maintained because condensation occurs at or near saturation given a sufficient supply of condensation nuclei.}{The key assumption in our theory is that the surface pressure is everywhere equal to the saturation vapor pressure of the atmospheric constituent corresponding to the local surface temperature. For a pure condensible atmosphere the partial pressure of the condensible substance is in fact the total pressure, so that the statement that a layer of the atmosphere is saturated is equivalent to the statement that the atmospheric pressure equals the saturation vapor pressure corresponding to the local atmospheric temperature; insofar as the temperature of the air in contact with the surface must equal the surface temperature, our assumption about surface pressure is equivalent to the statement that the atmosphere is saturated at the ground. More simply put, the atmosphere is assumed to be in gas/condensate equilbrium where it contacts the condensate reservoir at the ground. Where the atmosphere is cold enough to be condensing, it will be in gas/condensate equilibrium with the condensate particles that form {\it in situ} (assuming a sufficient supply of condensation nucleii), and the requirement for a pre-existing surface condensate reservoir can be relaxed. 

Although for the most part our theory only requires the atmosphere to be saturated at the surface, some assumption about the vertical temperature structure must be made in order to compute $OLR$ as a function of surface temperature. Assuming that the atmosphere is saturated throughout its depth determines the vertical temperature structure, and we will make use of this assumption in the $OLR$ calculations carried out in the examples we provide below. The presence of a subsaturated later in the high optically thin parts of the atmosphere would do little to change the results, but a deep subsaturated layer extending nearly to the surface -- such as could be produced by strong near-infrared heating of the atmosphere owing to instellation by an M-star -- would require a more complex approach to computing the $OLR$ on the dayside. 

If the hydrostatic surface pressure due to the atmosphere were significantly lower than the saturation vapor pressure corresponding to the surface temperature, then a layer of gas would form just above the surface having pressure equal to the saturation vapor pressure. There would be a pressure discontinuity between this layer and the lower overlying pressure, driving an extremely strong nonhydrostatic vertical flow that would add mass to the atmosphere, and simultaneously cool the surface through evaporation or sublimation, until the pressures are equalized.  Where the surface vapor pressure becomes lower than that of the overlying atmosphere, the same process happens in reverse, with mass and heat flowing onto the surface. Condensation may also occur interior to the atmosphere and not just at the surface, owing to radiative cooling inside the atmosphere. 
} 

The saturation assumption needs to be modified for the lava planet case, since mineral vapors in equilibrium with silicate melt are generally subsaturated relative to pure vapor condensation. \added{Vapor pressures for various mineral vapors in equilibrium with silicate melt are given in \citet{miguel2011compositions}, and at typical lava planet temperatures the vapor pressures for the dominant vapors (Na and SiO) are about an order of magnitude lower than the saturation vapor pressure in equilibrium with the respective pure liquids. }The lava planet case will be briefly discussed in Section \ref{sec:slow}. 

Since the surface temperature and pressure are so closely tied by the Clausius-Clapeyron relation for a saturated pure condensible atmosphere, the surface pressure gradient points from the cold region to the hot region in the pure condensible atmosphere. Therefore the atmosphere circulation is characterized by a strong low-level flow from warm region that carries both momentum and energy fluxes to cold places, and is referred to as ``sublimation-driven flow" in \citet{ingersoll_supersonic_1985}. In contrast, atmospheres with dilute condensible component are similar to dry atmospheres with surface pressure gradient pointing from the hot substellar or equatorial region to cooler regions, if the rotation effect of the planet is weak. Such atmospheres are characterized by an overturning circulation rising from the warm region and sinking in cool regions. The Hadley cells and Walker cell in the Earth's tropics are examples of this type of atmospheric circulation. In terms of mass balance, in dilute atmospheres, the convergence of the low-level flow is balanced by the divergence of the high-level return flow, while in pure condensible atmospheres, it is balanced by the local evaporation\added{/sublimation} minus precipitation flux through the surface. There is no pressure gradient that could support any return flow in the upper atmosphere. 

We start from the steady-state energy balance at the top of the atmosphere  (TOA).  For any atmosphere the net radiative flux at TOA must be balanced by the divergence of horizontal energy transport in the climate system including both in the atmosphere and the ocean (if present). Here we shall neglect oceanic transport. The atmospheric energy transport consists of transport of moist static energy plus transport of kinetic energy.  The former is composed of three terms: (1) potential energy flux; (2) sensible heat flux and (3) latent heat flux \citep{peixoto_physics_1992}. The specific potential energy, $gZ$ is typically of the same order as, but somewhat smaller than, the sensible heat $c_pT$. This can be seen by using the scale height $H \equiv {RT} / {g}$ as the estimate for $Z$, whence $({gZ})/({c_pT}) \approx {R}/{c_p}$, where $R$ is the gas constant for the atmospheric composition. For a pure condensible atmosphere, the specific latent heat of the atmosphere is the latent heat $L$ of the condensible gas, so the relative importance of sensible heat is given by the ratio ${c_p T}/{L} = {T}/{T^*}$, where $T^* \equiv {L}/{c_p}$.  $T^*$ is large for common condensibles.
For example, based on thermodynamic properties near the triple point of the respective gases,
it is 1350K for $\mathrm{H_2O}$, 484K for $\mathrm{CO_2}$, 210K for $\mathrm{N_2}$ and 3427K for $\mathrm{Na}$. All of these values are considerably supercritical in the sense of phase change \replaced{thermodynamcs}{thermodynamics}, so that for any pure-condensible atmosphere that operates in a substantially subcritical temperature range latent heat transport will dominate the sensible heat transport.

We will also neglect kinetic energy transport, which is the chief assumption by which our analysis departs from, and is much simpler than, the formulation of \citet{ingersoll_supersonic_1985}. If the speed of sound is used as a scale for the velocity, then the specific kinetic energy \replaced{$ v^2  /2  =  \gamma RT /2$}{$ v^2  /2$  equals to $ \gamma RT /2$}, which is of the same order as the specific sensible heat. Hence when latent heat dominates the sensible heat, it will also dominate the kinetic energy transport, except where the atmosphere becomes substantially supersonic.  Our approximate will therefore break down in supersonic regions of a very localized atmosphere with strong pressure gradients, but it nonetheless will suffice to discriminate between situations in which a local vs. global atmosphere occur. 

When the latent heat flux dominates the atmospheric energy transport,  the vertically integrated energy budget of the pure condensible atmosphere can be written as
\begin{equation} \label{eq:TOA}
F_\circledast (1-\alpha)\cos\zeta - OLR \simeq {\bm \nabla}_h \cdot (L {\bm M} )
\end{equation}
where the left hand side is the TOA radiative imbalance, and ${\bm M}$ the atmospheric horizontal mass transport vector. $OLR$ is the outgoing longwave radiation, $F_\circledast$ is the incident parallel-beam stellar flux, $\alpha$ the local planetary albedo and $\zeta$ the local zenith angle. \added{Since this is a vertically integrated equation, the divergence is in the horizontal direction alone.  The problem will be closed by writing $OLR$ as a function of surface temperature, but the statement remains valid regardless of how the $OLR$ is determined.}

In \citet{ingersoll_supersonic_1985} for the case of Io and \citet{castan_atmospheres_2011} for lava planets the energy carried by latent heat is neglected\deleted{, but because the atmosphere is assumed so thin that it has negligible effect on the surface temperature, the neglect does not lead to any significant error}. \added{However, in both calculations, it is assumed the atmosphere does not carry enough heat to affect the surface temperature. The surface temperature is assumed the same as it would be for a planet with no atmosphere, and the resulting temperature is used to drive the atmospheric mass flow.  In this situation, heat transport is just a diagnostic, so errors in heat transport do not affect the results as long as the heat transport is indeed small. } We will see shortly that even within our own framework, which includes latent heat transport, the conclusions of these two studies that the atmospheres are localized near the substellar point remain valid. 

We will assume that the outgoing longwave radiation (OLR) can be written as a function of the surface temperature ($T_s$) \footnote{This is a good assumption.  If the atmosphere is optically thin, then the OLR is the blackbody emission from the surface. If not, the temperature jump between the surface and the surface air is small for a pure condensible atmosphere so that the OLR can still be expressed as a function of $T_s$. However, for very thin atmospheres, the surface pressure could differ greatly from the saturation vapor pressure corresponding to the surface temperature, owing to limits on the evaporation\added{/sublimation} rate imposed by the speed of sound.} 
that is related with the surface pressure ($p_s$) by Clausius-Clapeyron relation. Then the energy budget equation (Eq.~\ref{eq:TOA}) can be simplified as an ordinary differential equation on the spatial distribution of the surface pressure ($p_s$) because the surface momentum equation provides additional relation between $p_s$ and the horizontal mass transport. 
In the next two sections, we will consider two specific orbital configurations of the planet, and further simplify the energy budget equation (Eq.~\ref{eq:TOA}) in the two different climate regimes to find how the surface pressure variation is controlled by the latent heat transport.   

\section{Slowly and Synchronously rotating  planets} \label{sec:slow}

For slowly and synchronously rotating planets, it is convenient to use the tidally locked coordinate system (see Appendix B in \citet{koll_deciphering_2015} for details) with tidally locked latitude $\theta_{TL} = 0$ at the terminator and $\theta_{TL} = \pi/2$ at the substellar point, because slow planetary rotation leads to zonally symmetric flows in the tidally locked coordinate. Then the energy budget equation and vertically averaged momentum equations are
\begin{eqnarray}
 \frac{1}{a\cos \theta_{TL}} \frac{\upd~~~~}{\upd \theta_{TL}} \left( \cos \theta_{TL} L v_s p_s/g  \right) & \simeq & 
 	\begin{cases} 
		F_a \sin \theta_{TL} - OLR (T_s),  & \theta \in [0, \pi/2]\  (\mathrm{dayside}) \\
		- OLR (T_s). &    \theta \in [-\pi/2, 0]\ (\mathrm{nightside}) \\
	\end{cases} \\
\frac{1}{a} \frac{\upd p_s~~}{\upd \theta_{TL}} = - \frac{\rho_s C_d | v_s | v_s}{H}
\end{eqnarray}
where $a$ is the radius of the planet, $g$ the surface gravity, \replaced{$F_a \equiv (1-\alpha)F\circledast$}{$F_a \equiv (1-\alpha)F_\circledast$} the absorbed stellar radiation at the substellar point, $C_d$ the neutral drag coefficient, \deleted{and} $H$ the local isothermal scale height $RT_s/g$\added{, $T_s$ the surface temperature, $p_s$ the surface pressure,  $\rho_s$  the surface air density and $v_s$ the velocity at the top of the friction layer.
Following \citet{ingersoll_supersonic_1985} we have assumed that the velocity is nearly independent of height over the depth which contains most of the mass of the atmosphere}.  
The form of the vertically averaged momentum balance is approximate, and assumes that the surface drag is distributed over the scale height.  
The planetary albedo is assumed uniform on the dayside and thus is included in  $F_a$. Given the relation between $T_s$ and $p_s$ by Clausius-Clapeyron relation, the spatial distribution of either surface temperature or pressure can be solved numerically using standard techniques for ordinary differential equations (we used the 4th-order Runge-Kutta method).  Calculations reported here were carried out with $C_d = .01$\added{, corresponding to a moderately rough surface with a roughness length of 20 cm  (\citet{pierrehumbert_principles_2010}, Eq. 6.20)}.

\begin{figure}[ht]
  \centering
  \vspace{-10pt}
  \includegraphics[width=\columnwidth]{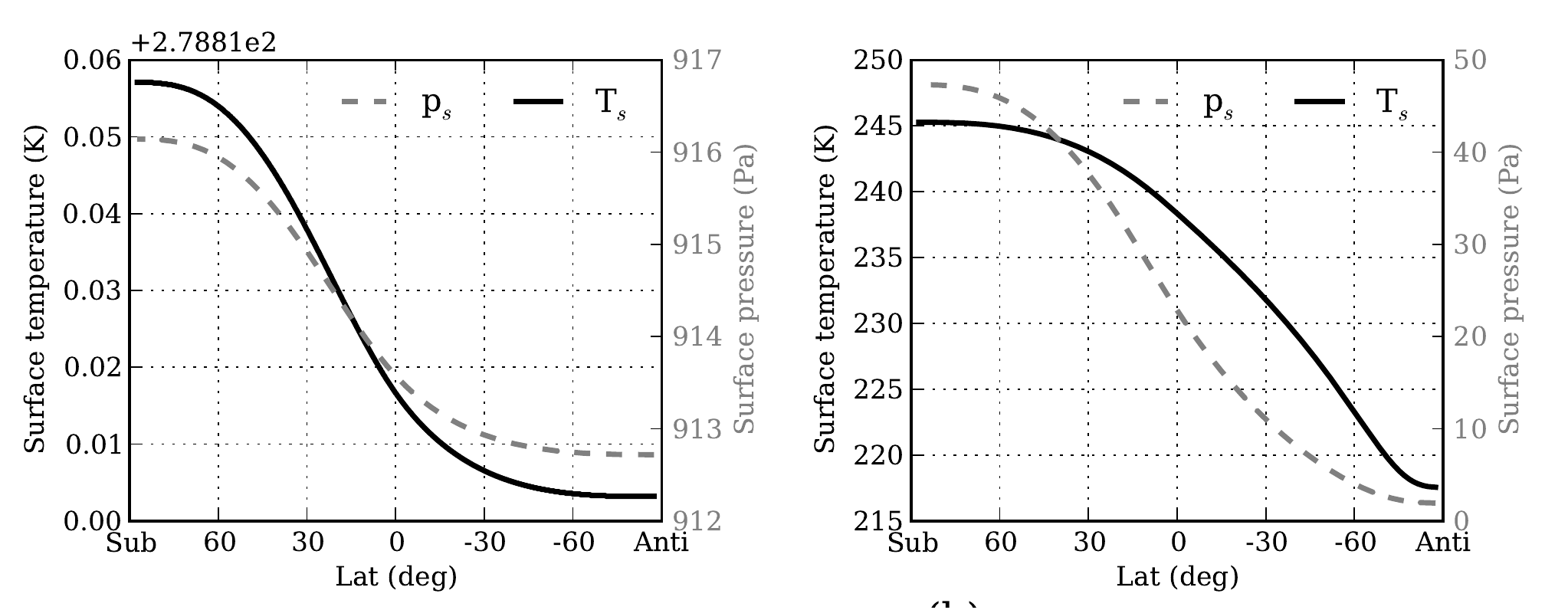}
  \caption{Meridional distribution of the surface temperature (solid black) and the surface pressure (gray dashed) in the tidally locked coordinate for the pure condensible H\2O atmosphere on a synchronously and slowly rotating planet, when (a) $F_a = 1112\,\mathrm{W\,m^{-2}}$ and (b) $F_a = 712\,\mathrm{W\,m^{-2}}$, respectively. Other parameters chosen to solve the 1D differential equation are: $a = r_\earth, g = g_\earth$. The OLR of a pure condensible H\2O atmosphere is fit by a third degree polynomial based on the realistic H\2O radiative transfer calculation: $OLR(T_s) = 4704.16 - 61.957T_s + 0.26912T_s^2-3.7244\times10^{-4}T_s^3$, when 200\,K$< T_s<$280\,K. }\label{fig:slow}
\end{figure}

Figure~\ref{fig:slow} shows an example of the solutions for \replaced{the}{a} pure condensible H\2O atmosphere. Solution in Figure~\ref{fig:slow}a under a higher insolation represents a warm and global H\2O atmosphere in an ocean world with surface pressure variation less than 1\%, while the one in Figure~\ref{fig:slow}b represents a cold local atmosphere concentrated on the day side of an icy planet with surface pressure variation of $\sim$95\%. These two solutions confirm that the atmospheric thickness dominates the latent heat transport that \replaced{flattens}{compensates for} the incoming differential radiative forcing, and plays an important role in the distribution of pure condensible atmospheres. The heat redistribution is still quite efficient when the substellar point is at the freezing point of water\deleted{, which precludes the proposed 'Eyeball Earth' state \citep{pierrehumbert_palette_2011} on a slowly and synchronously rotating planet. In other words, the planet with pure condensible H\2O atmosphere should be either a global ocean worlds or a global icy worlds}. \added{One consequence of this result is that, when the assumptions of our model are valid, Eyeball states with open water only near the substellar point such as discussed in \citet{pierrehumbert_palette_2011} can at best exist only in a very narrow range of instellations when water vapor is the only atmospheric constituent. Such worlds will tend to have surfaces which are either completely frozen over (in which case the atmosphere may be local if the system is cold enough) or completely liquid (in which case the atmosphere is global and the surface temperature is nearly uniform). The key difference with \citet{pierrehumbert_palette_2011} is that in the dilute case strong nightside temperature inversions can form which allow substantial geographic variations of surface temperature even when the free tropospheric temperature is fairly uniform. Because of the control of total surface pressure by Clausius-Clapeyron, such inversions cannot occur in the pure condensible
case, as they would be associated with extreme nonhydrostatic vertical pressure gradients that would drive a rapid flow to the surface, rectifying the imbalance quickly. Near the substellar point, the same processes operate in reverse, and keep the surface temperature from greatly exceeding the free tropospheric temperature at the top of the boundary layer. 
It remains to be seen how much noncondensible gas can be added to the atmosphere before strong nightside inversions and elevated substellar surface temperatures become possible again. }

We will further explore the effect of atmospheric thickness by considering the limit of global atmosphere with weak temperature variations $\Delta T \ll T_s$. 
In this limit, the surface pressure gradient could  still be large enough to maintain \replaced{the}{a} global atmosphere due to the Clausius-Clapeyron relation 
\begin{equation}
\frac{1}{p_s} \frac{\upd p_s}{\upd \theta_{TL}} = \frac{L}{R T_s} \frac{1}{T_s} \frac{\upd T_s}{\upd \theta_{TL}} 
\end{equation}
and the fact that $L/RT_s \gg 1$ for most condensing volatiles. 
Given the small surface temperature variation, $OLR \simeq F_a/4$\added{. Note that this constraint, which is a consequence of planetary energy balance, means that the behavior will be independent of the strength of the atmosphere's greenhouse effect, so long as our assumption that $OLR$ can be written as a function of $T_s$ is valid. We ignore the effect of clouds here which would alter the behavior both through their effect on $OLR$ and on the spatial distribution of absorbed stellar radiation.  However unlike in the Earth's tropics where clouds can be more variable than temperature, for pure condensible atmosphere buoyancy generation in the interior of atmosphere is prohibited and convection works very differently, in fact so differently it is unclear whether it should be called convection at all \citep{ding_convec_2016, pierrehumbert_dynamics_2016}. The cloud effect need to be evaluated consistently in 3D global climate models.
} \replaced{,   the}{The} fractional variation of $p_s$ then can be simplified as
\begin{equation} \label{eq:ps}
 \frac{1}{p_s} \frac{\upd p_s}{\upd \theta_{TL}}  \simeq  
 	\begin{cases} 
		\left( \mathcal{M} \frac{\sin \theta_{TL} + \cos 2\theta_{TL}}{\cos \theta_{TL}} \right) ^2,  & \theta \in [0, \pi/2]\  (\mathrm{dayside}) \\
		\left( \mathcal{M} \frac{\sin \theta_{TL} + 1}{\cos \theta_{TL}} \right)^2. &  \theta \in [-\pi/2, 0]\ (\mathrm{nightside}) \\
	\end{cases}
\end{equation}
where
\begin{equation}
\mathcal{M} = \frac{F_ag}{4Lp_s}      \frac{a \sqrt{a g C_d}}{R T_s} 
\label{eq:M}
\end{equation}
$\mathcal{M}$ is a non-dimensional parameter that can be used to distinguish whether the pure condensible atmosphere is global or local. Our assumption of a global atmosphere here requires that $\mathcal{M} \ll 1$. On the contrary, if $\mathcal{M} \geq O(1)$, the pure condensible atmosphere should exhibit large pressure variations and become local around the mass source. For example, consider the numerical solutions of the pure condensible  H\2O atmosphere in Figure~\ref{fig:slow}. For the global atmosphere solution in Figure~\ref{fig:slow}a, $\mathcal{M} = 0.046 \ll 1$; for the local atmosphere solution in Figure~\ref{fig:slow}b,  $\mathcal{M} = 1.46$. 

The non-dimensional parameter $\mathcal{M}$ can be understood by being rearranged into the ratio of two timescales: $\mathcal{M} = t_{adv} / t_{evap}$, where $t_{evap} = (p_s/g) / (F_a / 4L)$ is the characteristic time scale to build up  the condensible atmosphere by evaporation\added{/sublimation} under the global averaged insolation $F_a/4$, and $t_{adv} = a \sqrt{a g C_d} / (RT_s)$ the characteristic time scale to transport the air mass from the dayside to the nightside. Small values of $\mathcal{M}$ means that the atmospheric transport is fast enough to redistribute air mass from the hot place to other region and thus results in a global atmosphere, and vice versa. 

\citet{ingersoll_dynamics_1990} suggested a similar non-dimensional parameter to distinguish the pure condensible atmosphere, $\mathcal{M_I} = ({F_a g a / 4 p_s L}) / {c_s}$, where the numerator is the velocity scale at the terminator estimated by assuming half of the evaporated mass on the dayside should be transported to the nightside in a global pure condensible atmosphere, and the denominator the speed of sound corresponding to the global surface temperature. 
Our proposed  non-dimensional parameter  can also be rearranged in a  similar way, $\mathcal{M} = ({F_a g a / 4 p_s L}) / {v_s}$. The only difference is that the denominator here is the characteristic velocity scale if the atmosphere is local and the fractional variation of the surface pressure is $O(1)$ rather than the speed of sound, while the numerator is the same as the one defined in $\mathcal{M_I}$ -- the required mass transport to maintain a global pure condensible atmosphere .

\begin{figure}[ht]
  \centering
  \vspace{-10pt}
  \includegraphics[width=0.6\columnwidth]{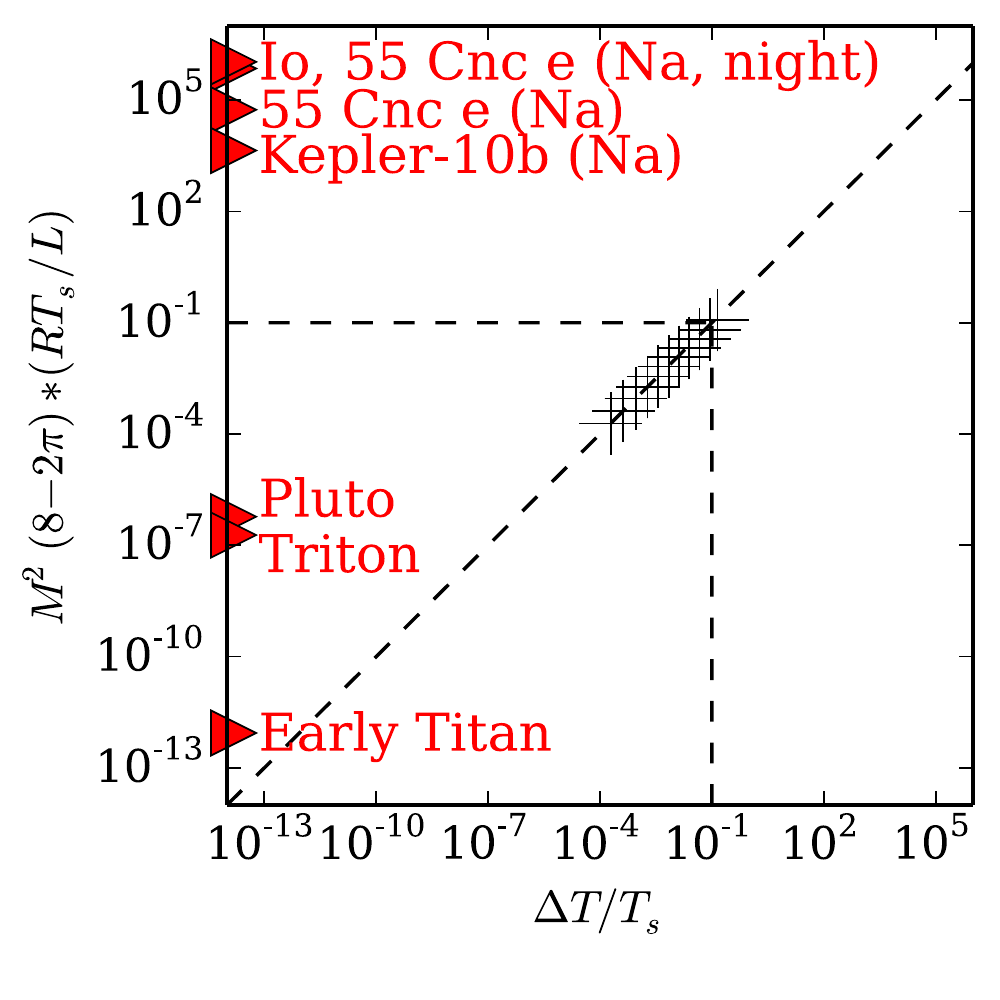}
  \caption{Theoretical fractional change of the surface temperature $\Delta T / T_s$ in a pure condensible as a function of the expression $\mathcal{M}^2 (8-2\pi) \frac{RT_s}{L}$.  Numerical results for a pure H\2O atmosphere based on $\Delta T$ from the substellar point to the antistellar point under various values of insolation between 712 and 1112\,$\mathrm{W\,m^{-2}}$, are shown as crosses. The non-dimensional expression $\mathcal{M}^2 (8-2\pi) \frac{RT_s}{L}$ of several planetary bodies with pure condensible atmospheres are estimated and marked on the vertical axis. For Io the atmosphere is assumed to be $\mathrm{SO_2}$, for Pluto, Triton and Early Titan $\mathrm{N_2}$, and for the lava planets Na.  The scaling theory is only quantitatively valid for small values of $\mathcal{M}$, but large values are sufficient to indicate regimes in which the atmosphere is very localized to the day-side of the planet. }\label{fig:dt}
\end{figure}

The non-dimensional parameter $\mathcal{M}$ can also be used to estimate the surface temperature variation ($\Delta T_s$) of the global pure condensible atmosphere by integrating Eq~(\ref{eq:ps}) from the substellar point to the antistellar point. The result is
\begin{equation}
\frac{T_{sub} - T_{anti}}{\bar{T_s}} \sim \mathcal{M}^2 (8-2\pi) \frac{RT_s}{L} 
\label{eqn:dtRelSlow}
\end{equation}
We calculate the numerical solutions of the fractional change of the surface temperature $\Delta T / T_s$ in the pure condensible H\2O atmosphere under various values of insolation between 712 and 1112\,$\mathrm{W\,m^{-2}}$, shown in Figure~\ref{fig:dt}. It is interesting that the expression in Eq. \ref{eqn:dtRelSlow} gives an accurate estimate of surface temperature variation not only for small values of $\mathcal{M}$ but also when $\mathcal{M}$ is order unity. We also show estimates of the the right hand side of Eq. \ref {eqn:dtRelSlow}  for several planetary bodies with  pure condensible atmospheres.   The estimates are consistent with either observations, or general circulation simulations published elsewhere: the condensing N\2 atmosphere on Triton, on Pluto \citep{forget2017Pluto}, and possibly on Early Titan \citep{charnay_titans_2014} should have small surface temperature variations; the condensing SO\2 atmosphere on Io should have large surface temperature variation as in \citet{ingersoll_supersonic_1985}. The magnitude of $\mathcal{M}$ on these planetary bodies are primarily determined by that of $p_s$ as discussed in \citet{trafton_global_1983} and \citet{trafton_large_1984}, and therefore is ultimately determined by the absorbed stellar radiation. Other parameters in $\mathcal{M}$ on these planetary bodies do not vary too much.   

A complete treatment of the lava planet case would require some modifications to our formulation, because vapor pressure of gases in equilibrium with a magma ocean are substantially subsaturated relative to the Clausius-Clapeyron relation for a pure gas in equilibrium with its liquid form (e.g. sodium vapor in equilibrium with a liquid sodium ocean).  A crude application of the basic model to the lava planet case is nonetheless sufficient to confirm that the atmosphere will be localized to the dayside and cannot carry sufficient heat to extend the magma ocean to the nightside.  We first do this by a form of {\it reductio ad absurdum}.  Suppose that the magma ocean extends globally. In that case, we can do a Clausius-Clapeyron fit to the sodium vapor pressure curve over a silicate melt, based on the data in \citet{miguel2011compositions}, which is valid wherever the vapor is in equilibrium with the silicate melt (hence valid globally if the magma ocean is assumed global).  With the modified effective Clausius-Clapeyron coefficients from this fit, our standard theory can be applied.  The controlling nondimensional parameter assuming a global magma ocean for the case of 55 Cancri-e and  Kepler 10b are shown in Figure \ref{fig:dt}, and show clearly that the atmospheres are in a similar regime to the localized atmosphere of Io, thus invalidating the assumption of a global magma ocean maintained by the evaporation-driven atmosphere. If the magma ocean is localized to the dayside, then the sodium vapor atmosphere must travel some distance past the edge of the magma ocean before it cools down enough to condense. As a crude indication of the degree of localization in this regime, we also plot in Figure \ref{fig:dt} the nondimensional parameter computed based on the Clausius-Clapeyron relation for sodium vapor, but using the temperature at which sodium at the vapor pressure at the edge of the magma ocean begins to condense (650K). The modified calculation also puts the atmosphere in the same localized regime as the Io's $\mathrm{SO_2}$ atmosphere. This calculation confirms that a thick noncondensible background atmosphere is required to account for the relatively high nightside temperature of 55 Cancri-e, as discussed in \citet{hammond2017Cancri}.

\section{Fast-rotating planets with zonally symmetric stellar forcing}\label{sec:fast}

On fast-rotating planets, Coriolis force becomes important in the atmospheric dynamics. We further simply the problem by assuming that the surface reservoir has a thermal inertia large enough to damp out the diurnal cycle, so the atmospheric flow is zonally symmetric in the geographic coordinate system. Then the meridional atmospheric transport that redistributes mass, momentum and energy is characterized by the Ekman transport in the frictional boundary layer, as discussed in \citet{pierrehumbert_dynamics_2016}. In the geographic coordinate system, the energy budget equation (Eq.\ref{eq:TOA}) becomes
\begin{equation} \label{eq:ekman_e}
 \frac{1}{a\cos \theta} \frac{\upd}{\upd \theta} \left( \cos \theta L M_E(\theta)  \right)  \simeq  
		F_a \cos \theta / \pi - OLR (T_s(\theta)). 
\end{equation}
where $\theta$ is the geographic latitude, $M_E(\theta)$ the meridional Ekman mass transport, and $F_a$  the absorbed stellar flux at the substellar point. Assuming that the frictional force in the Ekman layer is parameterized by a constant eddy viscosity $A$, the Ekman mass transport can be described by the product of the layer thickness and the zonal geostrophic  wind speed \citep[p.~112]{vallis_atmospheric_2006}, and thus is related to the meridional surface pressure gradient,
\begin{equation} \label{eq:ekman_m}
M_E \simeq - \frac{1}{4a\Omega} \sqrt{\frac{A}{\Omega}} (\sin^{-3/2} \theta )\frac{\upd p_s}{\upd \theta}  
\end{equation}
where $\Omega$ is the angular spin rate of the planet. Similar to the discussion about the synchronously and slowly rotating planets, the meridional distribution of either the surface temperature and pressure can be solved numerically, and  a non-dimensional parameter can be derived to distinguish whether the pure condensible atmosphere is global or local, $\mathcal{M_F} = \sqrt{\Omega/A} ( F_a a^2 \Omega) / (4 L p_s)$. Compared to the non-dimensional parameter defined in the synchronously and slowly rotating case, $\mathcal{M_F}$ has the same dependence on the insolation,  the specific latent heat and the surface pressure. However, $\mathcal{M_F}$ here is independent of the gravity because the Ekman transport has no dependence on the gravity. Similarly, $\mathcal{M_F}$ can also be rearranged as the ratio of the mass transport to maintain a global pure condensible atmosphere to the one if the atmosphere is local and the fractional variation of the surface pressure is $O(1)$. 

\begin{figure}[ht]
  \centering
  \vspace{-10pt}
  \includegraphics[width=0.6\columnwidth]{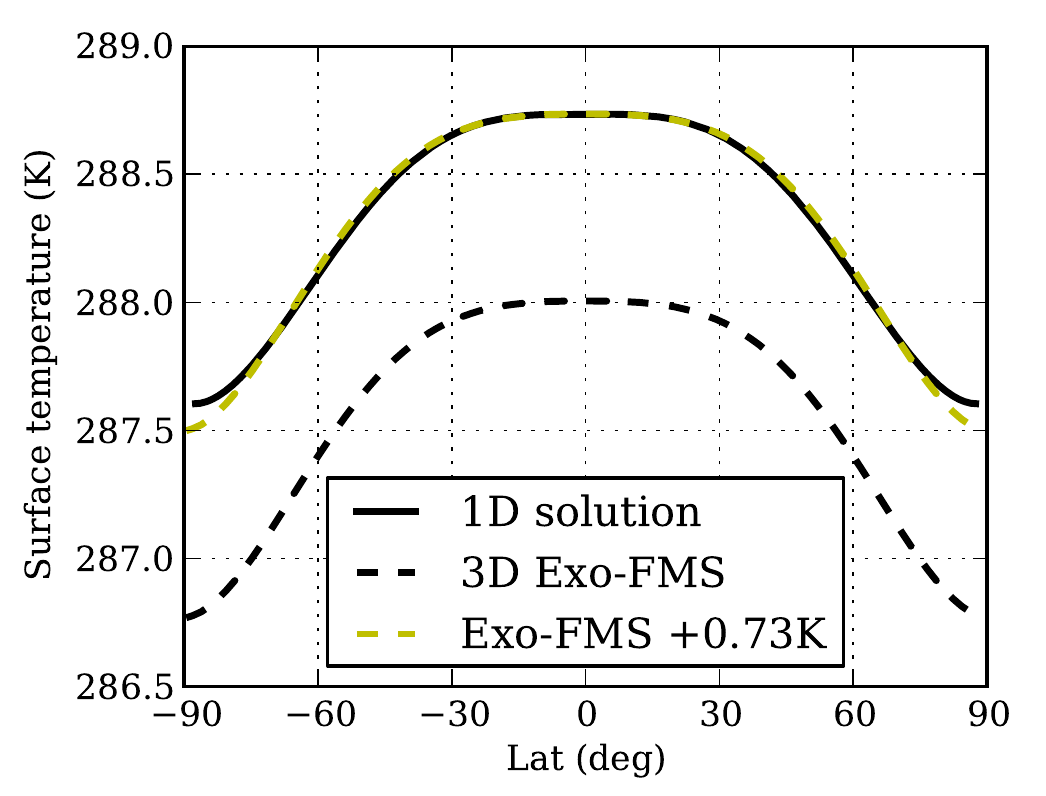}
  \caption{Meridional distribution of the surface temperature  for a pure condensible H\2O atmosphere on an Earth-like fast-rotating planet. The black solid  curve represents the solution of the 1D energy budget equation, and the black dashed  one the solution given by 3D general circulation model in \citet{pierrehumbert_dynamics_2016}. 
The 1D solution is generally 0.73\,K higher than the 3D solution due to the energy imbalance of the GCM, which is marked by the yellow dashed curve.
Parameters chosen to solve the 1D differential equation are: $a = r_\earth, g = g_\earth, F_a = 1260\,\mathrm{W\,m^{-2}}, A = 100\,\mathrm{m^2\,s^{-1}}$. The OLR of the pure condensible H\2O atmosphere is given by a linear fit based on the gray radiation scheme in the 3D GCM: $OLR(T_s) = 17.53 + 1.032T_s$. }\label{fig:fast}
\end{figure}

\citet{pierrehumbert_dynamics_2016} developed a 3D general circulation model to study the basic features of condensible-rich atmospheres, including the case of a global pure condensible H\2O atmosphere. \replaced{Tthe}{The} meridional distribution of $T_s$ from the 3D calculation is given by the black dashed curve in Figure~\ref{fig:fast}. We solved the 1D energy budget equation using the same input parameters and a typical value of the boundary layer eddy viscosity $A = 100\,\mathrm{m^2\,s^{-1}}$. The 1D solution (black solid curve in  Figure~\ref{fig:fast}) is  generally 0.73\,K higher than the 3D solution due to the energy imbalance of the GCM ($\sim 0.75\,\mathrm{W\,m^{-2}}$), but has the same curvature as the 3D solution. This result confirms that for the rapidly rotating pure condensible case, the simple 1D solution of the energy budget equation reproduces the result given by more comprehensive 3D models, and the 1D energy budget equation is a generally useful tool to study pure condensible atmospheres. 

\section{Discussion} \label{sec:conclusions}

We have used pure water vapor atmospheres in many of our examples, but the general principles discussed apply to any pure-condensible atmosphere.  The water vapor atmospheres themselves are of potential interest for very water-rich planets, but it is unclear if such atmospheres can exist except as transients, because the absence of cold-trapping generally leads to hydrogen escape and build-up of a noncondensible $\mathrm{O_2}$ atmosphere \citep{wordsworth_abiotic_2014}.  However, it is possible that massive planets (inhibiting escape) with a strong oxygen sink at the surface could maintain water-dominated atmospheres. Given uncertainties about such sinks, and also the possibility that a planet could be caught in a transient stage (including, perhaps, the early stages of a runaway greenhouse) it is worth knowing what the observational signatures of a water-dominated atmosphere would be. 


The pure water case is relevant to the hypothetical water dominated atmosphere considered by \citet{turbet2016habitability} in their series of simulations of the climate of Proxima Centauri b.  \replaced{This case was not absolutely pure water vapor, but the amount of non-condensible nitrogen and $\mathrm{CO_2}$ included is so low as to make the pure water case a reasonable approximation.  The 3D simulations shown in their Fig.~6 (lower panel) and Fig.~7 are consistent with our predictions in the sense that a global water vapor atmosphere is produced by the dayside open-water region, but differ from our model predictions in that the uniform temperature only applies in the free troposphere.  Near the surface, a boundary layer develops in which there are substantial temperature gradients as in the familiar case of a noncondensible-dominated atmosphere \citep{pierrehumbert_palette_2011,pierrehumbert_dynamics_2016}.  Surface temperature varies between  320\,K at the substellar point and 200\,K on the nightside.  Nightside temperature profiles are not shown, but the substellar profile ( Fig.~7 of \citet{turbet2016habitability} ) exhibits a thin sub-saturated boundary layer with surface pressure of 0.01 bars, which would lead to a pressure discontinuity of nearly 0.2 bar with the saturation vapor pressure of the surface itself.  In this situation, the enormous pressure gradient would drive rapid evaporation \citep{ingersoll_supersonic_1985}, cooling the surface and saturating the boundary layer.  We suggest that the surface latent heat flux and boundary layer scheme used in the simulations of \citet{turbet2016habitability} are inaccurate when applied to the nearly pure condensible case but the matter deserves further study.}{Our analysis predicts that for substellar temperatures in the vicinity of freezing point or warmer the surface temperature should be very nearly uniform in a pure water case. The most water-dominated simulation of \citet{turbet2016habitability} is radically different from this prediction, exhibiting surface temperatures in excess of 320K in a substellar ocean with the rest of the planet frozen over and nightside temperatures  as low as 215K, though the free tropospheric temperature is quite uniform globally.  This case includes 0.01 bar of noncondensing $\mathrm{N_2}$, and the specific humidity at the substellar point shown in their Fig.~7 is   about 70\%, and independent of altitude. The resulting atmosphere is highly unsaturated at the surface,with relative humidity only 7\% at the substellar point. This in itself violates one of the key assumptions of our analysis, and it would be tempting to conclude that the modest amount of noncondensible substance in the atmosphere has lead to a radically different behavior. We do not believe this to be the case; instead, it appears that problems with the formulation of the evaporative flux in  \citet{turbet2016habitability} have led to spurious subsaturation. Specifically, the saturation vapor pressure over water at 320K is 0.1 bars, and would lead to a layer of vapor immediately above the water surface with the same pressure. However, the hydrostatic pressure at the lowest model level (their Fig.~7) is an order of magnitude lower, at only 0.01 bar.  This should lead to an extremely strong evaporative flux of water vapor into the atmosphere -- in fact the upper layers of the ocean would literally boil into the atmosphere -- rapidly saturating the lower atmosphere and increasing atmospheric mass. Evidently, this process is not properly represented in the LMD model employed in \citet{turbet2016habitability}. The inconsistency we have identified depends only on the model surface pressure shown and the surface saturation vapor pressure, and is in no way affected by the presence of $\mathrm{N_2}$ in the atmosphere. 

The question remains, however, as to how the addition of a moderate amount of noncondensible gas would affect our results.  In addition, Proxima Centauri b has an orbital period of 11 days, so while Coriolis effects are weak they are far from negligible.  Both effects will require further study, most likely with the aid of 3D general circulation model simulations. }

In our own Solar System, the case of an evaporation\added{/sublimation}-driven pure water atmosphere may be relevant to one proposed scenario for the evolution of Pluto's crust. \citet{sekine2017charon} proposed that the reddish deposits of Cthulhu Regio near Pluto's equator could have been formed in melt-water pools resulting from melting crustal ice by the giant impact that formed Charon.  \replaced{Our calculation, however, indicates}{The scaling of temperature gradients given by Eqs.~\ref{eq:M} and \ref{eqn:dtRelSlow}, and illustrated by the warm case in Fig. \ref{fig:slow}, however, indicate} that the resulting water vapor atmosphere would likely become global and lead to global crustal melting, possibly diluting and globally redistributing organic material brought in by the impactor.  There is much to be done to explore this scenario, as it depends much on the lifetime of the transient water vapor atmosphere, but our calculations are sufficient to indicate that the possibility of a global ocean should be entertained.

The evaporation\added{/sublimation}-driven flow is a qualitatively different circulation regime as compared to the more familiar thermally direct overturning circulation. In the case of a local atmosphere, there would be a clear observational signal in the strong day/night contrast in conjunction with atmospheric spectral lines on the dayside detected by thermal emission. However, the case of a global pure-condensible atmosphere would generate a flat phase curve that could be hard to distinguish from that arising from other circulations with efficient horizontal energy transport.  For example, a thick atmosphere with a moderate amount of water vapor that is dominated by an overturning circulation could also generate the same signal \citep{yang_stabilizing_2013}. 

However, if a non-condensible radiatively-active substance exists as a minor constituent in \replaced{the global}{an otherwise} pure-condensible atmosphere, a unique observational signal could be detected. \replaced{As the consequence of the evaporation-driven flow in the pure condensible atmosphere, the non-condensible substance should accumulate near the antistellar point. This inhomogeneous distribution of the non-condensible substance induced by the atmospheric dynamics makes the thermal phase curve in the spectral region where the non-condensible substance absorbs reach the minimum at the antistellar point. In addition, the lack of absorption features of the non-condensible substance on the dayside detected by high-resolution spectroscopy could also imply such an inhomogeneous distribution.  We propose that  these observational features very likely indicate a global pure condensible atmosphere, if detected in the future.  By most currently employed methods, it is hard to constrain the bulk of the atmosphere and disentangle the degeneracy between the surface pressure and the molar concentrations from only the thermal phase curve.}{Suppose that the non-condensible gas has no sources or sinks, either at the surface or interior to the atmosphere. As a consequence of the evaporation/sublimation-driven flow in the pure condensible atmosphere, the non-condensible substance should accumulate near the antistellar point.
The resulting dayside/nightside asymmetry in concentration of the non-condensible substance would affect the spectrally resolved thermal emission phase curve of the planet.  Dayside emission before going into secondary eclipse would be lacking the absorption lines of the noncondensible gas.  Insofar as the terminator region would also be depleted in the noncondensible gas, a similar lack could be discerned in transit-depth spectroscopy.  Nightside thermal emission would on the other hand show a strong spectral signature of the noncondensing gas.  Such measurements for small, cool exoplanets  are probably beyond the reach of current technology, but it is to be hoped that future developments in hardware, as well as further refinements of such detection techniques as high-resolution spectroscopy, will make our suggested observational program feasible.} The technique we suggest here \replaced{can help us study}{could help to characterize} the atmospheric circulation in condensible-dominated atmospheres by making use of the distribution of a non-condensible minor constituent. \added{As an example of a related study that has been successful in the Solar System,} \citet{sprague_mars_2007} studied the atmospheric argon (Ar) \replaced{measurements}{concentration} in Mars' atmosphere from the gamma ray spectrometer on the Mars Odyssey spacecraft, and they found a significant enhancement of Ar  over the south polar latitudes occurring near the onset of the southern winter and rapid seasonal variations in Ar from 60$^\circ$S to 90$^\circ$S \replaced{that might be the evidence for wave activities, which provides an archetype for the kind of analysis we are suggesting. }{, owing to the concentration of Ar left behind when $\mathrm{CO_2}$ condenses out onto the surface.  This provides a valuable constraint on atmospheric transport mechanisms that are not  directly observable. While the enhancement of non-condensible gas on exoplanets would be detected by infrared spectroscopy rather than gamma ray spectroscopy, the general principles for interpreting the results are much the same. }

\acknowledgments
 
Support for this work was provided by the NASA Astrobiology Institutes Virtual Planetary Laboratory Lead Team, under the National Aeronautics and Space Administration solicitation NNH12ZDA002C and Cooperative Agreement Number NNA13AA93A, and by the European Research Council Advanced Grant EXOCONDENSE. 
 

\bibliography{con_rich}

\begin{thebibliography}{}
\expandafter\ifx\csname natexlab\endcsname\relax\def\natexlab#1{#1}\fi
\providecommand{\url}[1]{\href{#1}{#1}}

\bibitem[{Castan \& Menou(2011)}]{castan_atmospheres_2011}
Castan, T., \& Menou, K. 2011, ApJ, 743, L36.
\newblock \url{http://stacks.iop.org/2041-8205/743/i=2/a=L36}

\bibitem[{Charnay {et~al.}(2014)Charnay, Forget, Tobie, Sotin, \&
  Wordsworth}]{charnay_titans_2014}
Charnay, B., Forget, F., Tobie, G., Sotin, C., \& Wordsworth, R. 2014, Icarus,
  241, 269.
\newblock \url{http://linkinghub.elsevier.com/retrieve/pii/S0019103514003649}

\bibitem[{Ding \& Pierrehumbert(2016)}]{ding_convec_2016}
Ding, F., \& Pierrehumbert, R.~T. 2016, The Astrophysical Journal, 822, 24.
\newblock \url{http://stacks.iop.org/0004-637X/822/i=1/a=24}

\bibitem[{Forget {et~al.}(2017)Forget, Bertrand, Vangvichith, Leconte, Millour,
  \& Lellouch}]{forget2017Pluto}
Forget, F., Bertrand, T., Vangvichith, M., {et~al.} 2017, Icarus, 287, 54

\bibitem[{Forget {et~al.}(2013)Forget, Wordsworth, Millour, Madeleine, Kerber,
  Leconte, Marcq, \& Haberle}]{forget_3d_2013}
Forget, F., Wordsworth, R., Millour, E., {et~al.} 2013, Icarus, 222, 81.
\newblock \url{http://linkinghub.elsevier.com/retrieve/pii/S0019103512004265}

\bibitem[{Hammond \& Pierrehumbert(2017)}]{hammond2017Cancri}
Hammond, M., \& Pierrehumbert, R.~T. 2017, The Astrophysical Journal, 849, 152

\bibitem[{Ingersoll(1990)}]{ingersoll_dynamics_1990}
Ingersoll, A.~P. 1990, Nature, 344, 315.
\newblock \url{http://www.nature.com/doifinder/10.1038/344315a0}

\bibitem[{Ingersoll {et~al.}(1985)Ingersoll, Summers, \&
  Schlipf}]{ingersoll_supersonic_1985}
Ingersoll, A.~P., Summers, M.~E., \& Schlipf, S.~G. 1985, Icarus, 64, 375.
\newblock
  \url{http://www.sciencedirect.com/science/article/pii/0019103585900624}

\bibitem[{Koll \& Abbot(2015)}]{koll_deciphering_2015}
Koll, D. D.~B., \& Abbot, D.~S. 2015, The Astrophysical Journal, 802, 21.
\newblock
  \url{http://stacks.iop.org/0004-637X/802/i=1/a=21?key=crossref.68d921cd4e00b78e49f0486caf25900e}

\bibitem[{Miguel {et~al.}(2011)Miguel, Kaltenegger, Fegley, \&
  Schaefer}]{miguel2011compositions}
Miguel, Y., Kaltenegger, L., Fegley, B., \& Schaefer, L. 2011, The
  Astrophysical Journal Letters, 742, L19

\bibitem[{Peixoto \& Oort(1992)}]{peixoto_physics_1992}
Peixoto, J.~P., \& Oort, A.~H. 1992, Physics of {Climate} (American Institute
  of Physics)

\bibitem[{Pierrehumbert(2010)}]{pierrehumbert_principles_2010}
Pierrehumbert, R.~T. 2010, Principles of {Planetary} {Climate} (Cambridge
  University Press), google-Books-ID: bO\_U8f5pVR8C

\bibitem[{Pierrehumbert(2011)}]{pierrehumbert_palette_2011}
---. 2011, The Astrophysical Journal Letters, 726, L8.
\newblock \url{http://stacks.iop.org/2041-8205/726/i=1/a=L8}

\bibitem[{Pierrehumbert \& Ding(2016)}]{pierrehumbert_dynamics_2016}
Pierrehumbert, R.~T., \& Ding, F. 2016, Proceedings of the Royal Society A:
  Mathematical, Physical and Engineering Science, 472, 20160107.
\newblock
  \url{http://rspa.royalsocietypublishing.org/lookup/doi/10.1098/rspa.2016.0107}

\bibitem[{Sekine {et~al.}(2017)Sekine, Genda, Kamata, \&
  Funatsu}]{sekine2017charon}
Sekine, Y., Genda, H., Kamata, S., \& Funatsu, T. 2017, Nature Astronomy, 1,
  0031

\bibitem[{Sprague {et~al.}(2007)Sprague, Boynton, Kerry, Janes, Kelly, Crombie,
  Nelli, Murphy, Reedy, \& Metzger}]{sprague_mars_2007}
Sprague, A.~L., Boynton, W.~V., Kerry, K.~E., {et~al.} 2007, Journal of
  Geophysical Research: Planets, 112, doi:10.1029/2005JE002597, e03S02.
\newblock \url{http://dx.doi.org/10.1029/2005JE002597}

\bibitem[{Trafton(1984)}]{trafton_large_1984}
Trafton, L. 1984, Icarus, 58, 312.
\newblock \url{http://linkinghub.elsevier.com/retrieve/pii/0019103584900484}

\bibitem[{Trafton \& Stern(1983)}]{trafton_global_1983}
Trafton, L., \& Stern, S.~A. 1983, The Astrophysical Journal, 267, 872.
\newblock \url{http://adsabs.harvard.edu/doi/10.1086/160921}

\bibitem[{Turbet {et~al.}(2016)Turbet, Leconte, Selsis, Bolmont, Forget, Ribas,
  Raymond, \& Anglada-Escud{\'e}}]{turbet2016habitability}
Turbet, M., Leconte, J., Selsis, F., {et~al.} 2016, Astronomy \& Astrophysics,
  596, A112

\bibitem[{Vallis(2006)}]{vallis_atmospheric_2006}
Vallis, G.~K. 2006, Atmospheric and {Oceanic} {Fluid} {Dynamics}:
  {Fundamentals} and {Large}-scale {Circulation} (Cambridge University Press)

\bibitem[{Wordsworth \& Pierrehumbert(2014)}]{wordsworth_abiotic_2014}
Wordsworth, R., \& Pierrehumbert, R. 2014, The Astrophysical Journal Letters,
  785, L20.
\newblock \url{http://stacks.iop.org/2041-8205/785/i=2/a=L20}

\bibitem[{Wordsworth {et~al.}(2010)Wordsworth, Forget, Selsis, Madeleine,
  Millour, \& Eymet}]{wordsworth_is_2010}
Wordsworth, R.~D., Forget, F., Selsis, F., {et~al.} 2010, Astronomy \&
  Astrophysics, 522, A22.
\newblock \url{http://www.aanda.org/10.1051/0004-6361/201015053}

\bibitem[{Yang {et~al.}(2013)Yang, Cowan, \& Abbot}]{yang_stabilizing_2013}
Yang, J., Cowan, N.~B., \& Abbot, D.~S. 2013, ApJ, 771, L45.
\newblock \url{http://iopscience.iop.org/2041-8205/771/2/L45}

\end{thebibliography}
\bibliographystyle{aasjournal}
%



\listofchanges

\end{document}